\documentclass[letterpaper,10pt,oneside,twocolumn]{article}
\usepackage[margin=0.6in]{geometry}
\usepackage{array}
\usepackage[caption=false]{subfig}
\usepackage{stfloats}
\usepackage{url}
\usepackage{verbatim}
\usepackage{graphicx}
\usepackage{balance}

\usepackage{amsmath, amssymb, amsthm, amsfonts}
\usepackage{mathtools}
\usepackage{siunitx}
\usepackage[table, dvipsnames]{xcolor}

\usepackage{tikz}
\usetikzlibrary{calc, shapes, backgrounds, bending, decorations.pathmorphing, positioning, decorations.pathreplacing, shapes.geometric, graphs}
\usetikzlibrary{spy}
\usepackage{pgfplots}  
\pgfplotsset{compat=newest}
\usepgfplotslibrary{colorbrewer}
\usepgfplotslibrary{polar}
\usepgfplotslibrary{statistics}
\usepackage{booktabs}
\usepackage{bm}
\usepackage{cite}
\DeclareSIUnit\radian{rad}
\usepackage{algorithm}
\newcommand*{\tran}{^{\mkern-1.5mu\mathsf{T}}}
\usepackage[colorlinks=true, allcolors=blue]{hyperref}
\usepackage[nameinlink,capitalise]{cleveref}
\usepackage{authblk}

\begin{document}
\pgfplotsset{every axis plot/.append style={}}%
\def\largeplot{0.2\linewidth}%
\def\mainLineWidth{1.5pt}%
\def\refLineWidth{1pt}%
\def\layersep{2cm}%
\definecolor{mycolornored}{HTML}{e41a1c}%
\definecolor{mycolorfnnh}{HTML}{377eb8}%
\definecolor{mycolorshred}{HTML}{984ea3}%
\definecolor{mycolorfnn}{HTML}{ff7f00}%
\definecolor{mycolorplstm}{HTML}{f781bf}%
\definecolor{mycolorcell}{HTML}{4daf4a}%
\definecolor{mycolorae}{HTML}{a65628}%
\pgfplotsset{
	Mylinenored/.style={
		line width = \mainLineWidth,
		mark=x,
		unbounded coords=jump,
		color=mycolornored
	}
}%
\pgfplotsset{
	Mylinefnnh/.style={
		line width = \mainLineWidth,
		mark=+,
		unbounded coords=jump,
		color=mycolorfnnh
	}
}%
\pgfplotsset{
	Mylineshred/.style={
		line width = \mainLineWidth,
		mark=o,
		unbounded coords=jump,
		color=mycolorshred
	}
}%
\pgfplotsset{
	Mylinefnn/.style={
		line width = \mainLineWidth,
		mark=star,
		unbounded coords=jump,
		color=mycolorfnn
	}
}%
\pgfplotsset{
	Mylineplstm/.style={
		line width = \mainLineWidth,
		mark=triangle,
		unbounded coords=jump,
		color=mycolorplstm
	}
}%
\pgfplotsset{
	Mylinecell/.style={
		line width = \mainLineWidth,
		mark=diamond,
		unbounded coords=jump,
		color=mycolorcell
	}
}%
\pgfplotsset{
	Mylineae/.style={
		line width = \mainLineWidth,
		mark=square,
		unbounded coords=jump,
		color=mycolorae
	}
}%

\pgfplotsset{
	MyTrajectory/.style={
		width=300pt,
		height=300pt,
		xlabel={$x$ in \si{\meter}},
		ylabel={$y$ in \si{\meter}},
	}
}%
\title{Communication Reduction via\\ Semantic-Based Encoding in DMPC Using LSTMs}
\author[1]{Torben Schiz}
\author[2]{Pedro H.\ J.\ Nardelli}
\author[1]{Henrik Ebel}
\affil[1]{Dynamics \& Control, Dept.\ of Mechanical Engineering, LUT University, Lappeenranta, Finland}
\affil[2]{Cyber-Physical Systems, Dept.\ of Electrical Engineering, LUT University, Lappeenranta, Finland}
\date{}

\maketitle

\begin{abstract}
The communication demands of distributed model prediction control~(DMPC) can overwhelm even advanced wireless communication technologies as agents must exchange a significant amount of information at least once per time step.
To semantically reduce communication demands, this work employs encoder-decoder networks built around long-short term memory~(LSTM) cells in a distributed optimization algorithm.
Agents publish a reduced representation of a message and receivers reconstruct the original message upon reception.
In tests with reduced communication using formations of mobile robots, trained networks retain satisfactory performance and work reliably under conditions overwhelming full communication. 
As the results show, the usage of LSTMs either allows unprecedented reconstruction accuracy or the usage of different prediction-horizon lengths without the necessity to retrain. 
\end{abstract}

\noindent\textbf{Keywords:} Distributed MPC, communication reduction, LSTM, distributed control, semantic communication, multi-agent systems, robotics
\section{Introduction}
Networked control has long been heralded to be more robust, adaptable, and scalable than centralized control methodologies and better performing than decoupled ones. 
A timely application area are especially groups of collaborating robots, as  ground-based and air-based robots have made headlines and created disruption in sectors like logistics, security, and defense. 
Still, in robotics, the vast majority of present-day real-world applications, even if advanced automation is present, typically do not include autonomous swarm-like collaborative behavior and self-reliant multi-robot collaboration. 
Apart from application-end engineering challenges, some of the reasons for this may also be found in the demands of control methods proposed to achieve such behavior. 
In particular, and in sole focus of this work, model predictive control~(MPC), which has become widely popular in industrial applications~\cite{SamadEtAl20}, holds big promises in its distributed variants also for networked and collaborative control applications.
Distributed variants can also benefit from MPC's general advantages like an easy-to-define and easy-to-alter control goal, inclusion of (potentially collaborative) constraints, direct consideration of nonlinear dynamics, and naturally optimized performance. 
Many such distributed model predictive control~(DMPC) schemes and algorithms have been proposed, tailored to different control-problem setups and often with the aim to prove theoretical guarantees for the closed-loop behavior~\cite{NegenbornMaestre14, RawlingsMayneDiehl20, StewartWrightRawlings11, StombergEtAl25a, MullerRebleAllgower12, KohlerMullerAllgower19}. 
In DMPC, each agent (e.g., each mobile robot) makes its own decisions by solving its own optimal control problem, removing any single point of failure represented by a centralized decision maker. 
However, each agent's OCP considers information from those other agents that impact the optimality of its own decisions through couplings in the control goals (cost function), the constraints, and/or the dynamics. 
Such agents are called neighbors subsequently, and, in mobile robotics, the information exchange typically needs to happen explicitly via wireless communication.

In stark contrast to MPC, to the best knowledge of the authors, real-world applications of DMPC are presently lacking, especially in robotics, where the latter is in focus of this article due to the many worthwhile applications it presents to networked control.
One reason might be, as for the application of MPC, a lack of computation power to solve the underlying optimal control problems~(OCPs) in a timely-enough manner, especially considering that the OCPs then typically need to be solved on on-board computers tracking behind personal and industrial computers in performance. 
Indeed, earlier works see limited computation resources as an inhibitor for DMPC applications~\cite{VanParysPipeleers17}. 
However, the progress in mobile compute performance has been swift, and a recent work finds that, actually, ``communication is the bottleneck''~\cite{StombergEtAl25} in real-world DMPC applications in robotics where agent-to-agent communication happens wirelessly and computation takes place on board of each robot. 
This may seem surprising as also communication technology has experienced progress. 
However, common communication technologies are typically designed to be as application agnostic as possible, aiming at the reliable and timely delivery of messages independently of their relevance to the task. 
Semantic communications, including task-oriented and goal-oriented approaches,
instead assign value to information according to its significance for the decision
or actuation process performed at the receiver~\cite{Pappas2024Goal,Fountoulakis2023Goal}. 
More generally, semantic communications incorporate the application objective into
the communication process, enabling task-relevant information to be represented and
transmitted using fewer communication resources. 
This gives rise to semantic encoding and decoding mechanisms that can reduce network
traffic and facilitate timely delivery, although these gains may require increased
computational complexity at the transmitter and receiver~\cite{Chaccour2025LessData}. 
This perspective is particularly suitable for networked control, where communication,
computation, and control performance are intrinsically coupled~\cite{Silva2024Semantic}, and where the value of information and requirements emanating from those for communication might be vastly different than for more common use-cases of present communication technology as, e.g., represented by typical usage patterns of smartphones or personal computers. 
In this context, semantic communications can reduce the communication burden by
prioritizing compact, task-relevant information, thereby improving scalability and
enabling informative messages to be delivered within the timing constraints of the
control system
\cite{Uysal2022Semantic,Charalambous2026Goal}.

In this context, this article shows that the (communication-related) performance of DMPC can be vastly improved by a semantics-based encoding of the communication information on application level, without touching lower-level elements of the communication stack, preserving straight-forward applicability using standard communication technologies and protocols. 
To that end, the article proposes a machine-learning-based reduction of to-be-communicated data to a latent space at the transmitting agents prior to communication paired with a decoder at the receiving agents, which reconstructs the original message.
Novelly, one of the proposed approaches almost perfectly recovers the closed-loop performance of lossless, uncompressed communication while being much more robust to real-world imperfections of communication. Further, we propose approaches that need not be retrained every time a different prediction horizon is picked, trading some of the performance for generality.
These approaches are based on encoder-decoder recurrent neural networks~\cite{ChoEtAl14}, which can be trained for arbitrary sequence lengths in a single training run.

Commonly in robotics, there is a simulation-to-reality gap emanating from the complexity of the real system dynamics. 
However, there exist many successful experiments involving simple mobile robots, and models can capture the underlying physics well, making it possible to switch from simulation to experiment with relative ease in certain applications~\cite{FuchsEtAl26,Ebel21}.
However, in DMPC for mobile robots, there is a significant, additional sim-to-real gap introduced by communication, as the communication is often simplified in experiments by solving the OCP on external compute units, for example in~\cite{BurkVolzGraichen21, Ebel21, RosenfelderEbelEberhard22, StombergEtAl23, EbelRosenfelderEberhard24, GrafeEtAl25}, thus, bypassing wireless inter-agent communication.

In control literature, to the best of the authors' knowledge, only few works attempt to reduce the communication demands by reducing the message sizes in DMPC, e.g., for collaborating mobile robots, and research streams are largely separated from research on communications technology~\cite{LuoChenGuo22}.
A non-data-driven approach to reducing the communication in DMPC in robotics is employed in~\cite{WengleVaragnolo24} and~\cite{DiLeoAbadPregene17}, whereas~\cite{El-FerikSiddiquiLewis16} and~\cite{SchizEbel25} reduce the communication using data-driven methods.
Of the data-driven approaches, \cite{El-FerikSiddiquiLewis16} communicates the parameters of a neural network that is re-trained in every step prior to communication leading to prohibitively long computation times.
The inter-agent communication is reduced using an autoencoder in~\cite{SchizEbel25}, where, prior to communication, the to-communicate information is passed through the (semantic) encoder and after reception through the (semantic) decoder part, reconstructing the original message.
With this approach, simulated robot formations successfully completed tasks under communication conditions where the uncompressed communication failed.
No other neural network architectures were tested and compared to the autoencoder.
Additionally, as a standard autoencoder is trained for fixed input and output layer sizes, the approach from~\cite{SchizEbel25} requires retraining whenever the horizon lengths is changed.
In other words, for each horizon length a separate autoencoder is needed, complicating controller tuning as the prediction horizon is often seen as a key tuning parameter in (D)MPC. 
Thus, the state of the art leaves two key research questions.
Firstly, the question arises whether other neural-network architectures can be trained to achieve better results for a fixed prediction horizon length.
Secondly, one may ask how architectures that generalize to a variable horizon length can be furnished, removing the need for explicit retraining whenever the prediction horizon length is changed.
Both questions are attended in this work as follows.
\cref{sec:problemsetting} introduces a general DMPC optimal control problem and adapts it subsequently for formations of nonholonomic robot formations for asymptotically stabilizing setpoint tasks. The models used for communication reduction are introduced in~\cref{sec:nnarchitectures}, followed by a description of the training process in~\cref{sec:datageneration}.
The trained models are first validated in simulation to compare their respective performances when used in the optimization algorithm under ideal communication conditions (\cref{sec:simulativeresults}) before testing the communication reduction models under realistic conditions on embedded hardware (\cref{sec:numexperiments}).
Finally, \cref{sec:conclusion} summarizes the findings.

\section{Problem Setting}\label{sec:problemsetting}
\noindent This section first provides a general distributed MPC setup in~\cref{sec:ageneralsetup} highlighting that the proposed communication reduction methods can be applied to a wide class of cooperative control problems. 
Then, \cref{sec:modelproblem} shows how DMPC can be applied to formation control of mobile robots as the exemplary application of used in this work.

\subsection{General Setup}\label{sec:ageneralsetup}
\noindent This work considers a quite optimal control problem furnished as follows. A cost function $J(\bm z(t), \bm u(\cdot \,\vert \, t )) = \sum_{k=0}^{H-1} \ell (\bm z(t+k\,\vert\, t), \bm u(t+k\,\vert\, t))$ couples $N\geq2$ dynamically decoupled systems by defining a common goal.
In the cost function, $\ell\!:\mathbb R^{n_z} \times \mathcal U \to \mathbb R$ is the so-called stage cost and the length of the prediction horizon is $H\in\mathbb N$.
The states and inputs are given by $\bm z\in\mathbb R^{n_z}$ and $\bm u\in\mathbb R^{n_u}$, respectively.
The inputs are subject to independent, per-system constraints, constraining the concatenated inputs to the set $\mathcal U = \mathcal U_1 \times \cdots \times \mathcal U_N$.
The notation $(\cdot\,\vert\,t)$ marks a whole discrete-time trajectory along the prediction horizon as predicted at time step $t$, of which $(k\,\vert\,t)$ denotes the specific prediction for time step~$k$ where $k\in \mathbb Z_{t:t+H-1}$ with $\mathbb Z_{a:b}\coloneqq\lbrace a,\dots, b\rbrace$.
With this, the control task can be formulated in the form of the OCP
\begin{align}
\underset{\bm u(\cdot \,\vert \, t )}{\textnormal{minimize}} & \;\;  J(\bm z(t), \bm u(\cdot \,\vert \, t ))\label{eq:mpccost}\\
\textnormal{subject to} & \;\; \bm z(t+k+1\,\vert\, t) \nonumber \\
& \;\;\;\;= \bm f(\bm z(t+k\,\vert\, t), \bm u(t+k\,\vert\, t)), \label{eq:discretizationofsysdym}\\
& \;\; \bm u(t+k\,\vert\, t) \in \mathcal{U},\, k \in \mathbb Z_{0:H-1},\\
& \;\; \bm z(t\,\vert\, t) = \bm z(t), \label{eq:mpcinit}
\end{align}
which we aim to solve distributedly in the typical receding-horizon manner of MPC.
Further, \cref{eq:discretizationofsysdym} is the appropriately discretized dynamics of the multi-agent system, where, in this article's case, the dynamics of the involved robots are decoupled, meaning that they can be simply concatenated. 

\subsection{Model Problem}\label{sec:modelproblem}
\begin{figure}[btp]
    \centering
    \includegraphics[width=0.5\linewidth]{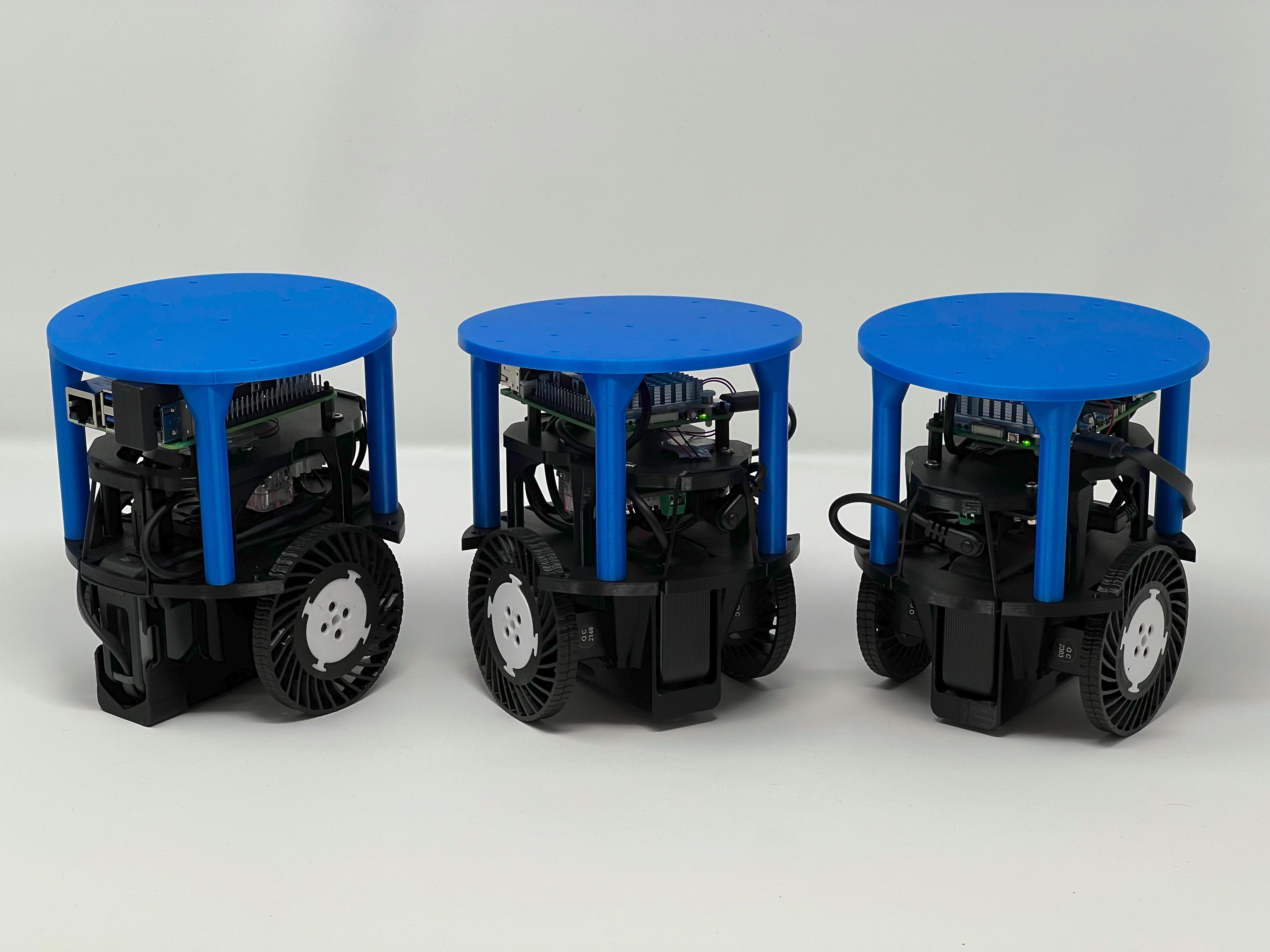}
    \caption{Photograph of differential-drive mobile robots as a common example for nonholonomic mobile robots}
    \label{fig:robexample}
\end{figure}

\noindent In the following, nonholonomic mobile robots, as shown in~\cref{fig:robexample}, are used as test systems, although the methods developed in this article do not explicitly depend on the very concrete control task and are adjusted to it only via observation data during training.
The continuous system dynamics of a single nonholonomic robot moving in a two-dimensional plane can be modeled using first-order kinematics as
\begin{align}
    \dot{\bm z}_i(t) = \begin{bmatrix}
        \dot x_i (t) \\ \dot y_i (t) \\ \dot\theta_i (t) 
    \end{bmatrix} = \begin{bmatrix}
        v_i(t) \cos(\theta_i(t)) \\ v_i(t) \sin(\theta_i(t)) \\ \omega_i (t)
    \end{bmatrix}
\end{align}
with the state vector ${\bm z}_i(t) = \begin{bmatrix}x_i (t) &  y_i (t) & \theta_i (t)\end{bmatrix}\tran \in \mathbb R^{n_i}$, $n_i=3$ and the inputs $\bm u_i(t) = \begin{bmatrix}
    v_i(t) & \omega_i(t)\end{bmatrix}\tran \in \mathbb R^{m_i}$, $m_i=2$.
The state vector consists of the position in the plane $\begin{bmatrix}
 x_i(t) & y_i(t)   
\end{bmatrix}\tran$ and the orientation $\theta_i(t)$ measured relative to the $x$ axis.
The input is made up of the linear velocity $v_i(t)$ and the angular velocity $\omega_i(t)$.
The nonholonomic differential-drive robot has two parallel wheels as well as a castor wheel or ball.
The implicitly fulfilled Pfaffian constraint $\begin{bmatrix}\sin \theta_i & -\cos\theta_i & 0\end{bmatrix}\dot{\bm z}_i = 0$ on the kinematics prevents instantaneous lateral motion of the robot.

In the following, a brief description of the concrete exemplary formation-control task for nonholonomic robots as solved in this work's numerical experiments is given. The task has previously been studied in~\cite{RosenfelderEbelEberhard22}, where a more detailed derivation can be found.
The overall system dynamics of the decoupled system results from concatenating the states and inputs  of the individual robots to
\begin{align}
    \dot{\bm z} (t) = \begin{bmatrix}
        \dot{\bm z}_1\tran (t) & \cdots & \dot{\bm z}_N\tran (t)   
    \end{bmatrix}\tran, \; \bm z(0) = \bm z_0.
\end{align}
The discretization in time with zero-order hold on the inputs and with sampling time $\Delta t$ is denoted as $\bm z(t + \Delta t) = \bm G_{\textnormal{d}} (\bm z(t), \bm u(t))$.
 
The output $\bm y^\mathrm{R}\in \mathbb R^{3(N+1)}$, describing the robot formation, is expressed through the relation $\bm y^\mathrm{R}(\bm z, \hat\theta_\mathrm{d}) = \bm C^\mathrm{R}(\hat\theta_\mathrm{d}) \bm z$ where $\hat\theta_\mathrm{d}$ is the desired orientation of the geometric center of the formation and the superscript $\bullet^\mathrm{R}$ is used to describe quantities expressed in the auxiliary reference frame $\mathcal K_\mathrm R$.
The auxiliary reference frame~$\mathcal K_\mathrm R$ arises by rotation of the inertial frame of reference by $\hat \theta_\mathrm{d}$.
Here and in the following, $\hat\bullet$ denotes a variable related to the geometric center of the formation.
Mapping the concatenated state vector to the output via the matrix $\bm C^\mathrm{R}\in\mathbb R^{3(N+1)\times 3N}$ adds the geometric center as the average of the poses of the robots in the formation, expresses the positions of each robot relative to said geometric center, and describes the positions in an auxiliary frame of reference~$\mathcal K_\textrm{R}$.
The orientation of each robot and the geometric center remain given in the inertial frame of reference.
Describing the positions of the robots relative to the geometric center allows penalizing the absolute position of the geometric center and the relative positioning within the formation individually.
To ensure that all robots and the geometric center can approach the desired setpoint along the $x^\mathrm{R}$-axis, only such scenarios are considered where all robots in the formation have identical desired orientations.
Adding all robots in the formation and the geometric center to the output introduces a redundancy to the formulation, which needs to be accounted for when picking a specific desired output
\begin{align}
    \bm y_\textnormal{d}^\textnormal{R} = \left[\begin{matrix}\hat x_\textnormal{d}^\textnormal{R} & \hat y_\textnormal{d}^\textnormal{R} & \hat \theta_\textnormal{d} & x_{\hat {\bm z} \to 1,\textnormal{d}} & y_{\hat {\bm z} \to 1,\textnormal{d}} & \theta_{1,\textnormal{d}} & \cdots \end{matrix}\right.\nonumber\\
    \left.\begin{matrix} \cdots &  x_{\hat {\bm z} \to N,\textnormal{d}} & y_{\hat {\bm z} \to N,\textnormal{d}} & \theta_{N,\textnormal{d}}  \end{matrix}\right]\tran \in \mathbb R^{3(N+1)}\,,
\end{align}
with $\theta_{j,\textnormal{d}} = \hat \theta_{\textnormal{d}}$ and where the subscript ${\hat {\bm z} \to j,\textnormal{d}}$ denotes a value as being relative to the geometric center for all $j \in \mathcal N$ with $\mathcal N$ the number of agents in the formation.
The cost function coupling the control systems and defining the common goal for each robot is chosen as
\begin{align}\label{eq:formation_cost}
    & J\!\left(\bm z(t), \bm u(\cdot \, \vert \, t), \bm y_{\textnormal{d}}^\textnormal{R}\right)  \nonumber \\
    & \;\; = \sum_{k = 0}^{H - 1} \ell\left(\bm z(t + k \,\vert\, t), \bm u(t + k \,\vert\, t) ,\bm y_{\textnormal{d}}^\textnormal{R}\right).
\end{align}
defined via the stage cost
\begin{align}\label{eq:formation_stage_cost}
    \ell (\bm z, \bm u, \bm y_{\textnormal{d}}^\textnormal{R}) &= \hat\ell (\bm z, \bm u, \bm y_{\textnormal{d}}^\textnormal{R})\nonumber \\
    &\;\;+ \sum_{j = 1}^N \left(\ell_{j,\textnormal{rel}} \left(\bm z, \bm u, \bm y_{\textnormal{d}}^\textnormal{R}\right) + \ell_{j,u} (\bm u)\right) \,,
\end{align}
composed of
\begin{align}
    \hat \ell (\cdot) &\coloneqq \hat d_1\left(\hat x^\textnormal{R} - \hat x_{\textnormal{d}}^\textnormal{R}\right)^4 + \hat d_2\left(\hat y^\textnormal{R} - \hat y_{\textnormal{d}}^\textnormal{R}\right)^2 \\ 
    &\;\;+ \hat d_3\left(\hat \theta^\textnormal{R} - \hat \theta_{\textnormal{d}}^\textnormal{R}\right)^4, \\
    \hat \ell_{j,\text{rel}} (\cdot) &\coloneqq \hat d_{1,j}\left(\hat x_{\hat {\bm z} \to j}^\textnormal{R} - \hat x_{\hat {\bm z} \to j,{\textnormal{d}}}^\textnormal{R}\right)^4 + \hat d_{2, j}\left(\hat y_{\hat {\bm z} \to j}^\textnormal{R} - \hat y_{\hat {\bm z} \to j,{\textnormal{d}}}^\textnormal{R}\right)^2 \nonumber \\
    &\;\; + \hat d_{3, j}\left(\hat \theta_j^\textnormal{R} - \hat \theta_{j,{\textnormal{d}}}^\textnormal{R}\right)^4, \\
    \ell_{j,u} (\cdot) &\coloneqq r_{1,j} v_j^4 + r_{2,j} \omega_j^4
\end{align}
with the weights $\hat d_i, \, d_{i,j}, \, r_{m,j} \in \mathbb R_{>0}$, $i \in \mathbb Z_{1:3}$, $j \in \mathcal N$, $m \in \mathbb Z_{1:2}$.
The weights are chosen for $j\in \mathcal N$ as $\hat d_1 = d_{1,j} = 1$, $\hat d_2 = d_{2,j} = 5$, $\hat d_3 = d_{3,j} = 0.1$, $r_{1,j} = 0.125$, and $r_{2,j} = 0.0125$ based on~\cite{worthmann2015regulation}.
The stage cost uses both quadratic and quartic terms as described in~\cite{RosenfelderEbelEberhard22} following~\cite{worthmann2015regulation} as even a single nonholonomic robot cannot be asymptotically stabilized to a full-state setpoint using merely quadratic terms when no terminal ingredients are used.
Combining the previously introduced ingredients yields the distributed optimal control problem for robot $i\leq N$ at time $t$ as
\begin{align}
    \underset{\bm u_{i}(\cdot \, \vert \, t)}{\text{minimize}} &\; \; J\!\left(\bm z(t), \bm u(\cdot \, \vert \, t), \bm y_\textnormal{d}^\textnormal{R}\right)\label{eq:dist_cost} \\
    \text{subject to} &\; \; \bm z(t + k + 1 \,\vert\, t) = \nonumber\\
& \; \;\; \; \bm G_\textnormal{d} \left(\bm z(t + k\,\vert \, t), \bm u(t + k \,\vert \, t)\right), \label{eq:dist_sys_dyn}  \\
    &\; \; \bm u(t + k \,\vert\, t) \in \mathcal U, \, k \in \mathbb Z_{0:H-1}, \label{eq:inputconstraints} \\
    & \; \;\bm z(t \,\vert\, t) = \bm z(t). \label{eq:initial_state_is_current}
\end{align}
Each robot optimizes only for its own control inputs $\bm u_i$ and not for the control inputs of the neighboring robots $j\in \mathcal N\backslash\lbrace i\rbrace$.
Inputs are constrained using box constraints $\mathcal U = \left[\underline{v}, \overline{v}\right] \times \left[\underline{\omega}, \overline{\omega}\right]\subset \mathbb R^{2}$ with $\underline{v} \leq \overline{v}$ and $\underline{\omega} \leq \overline{\omega}$.

To solve the distributed OCP, the robots in the formation use the distributed optimization algorithm from~\cite{StewartWrightRawlings11}, which has been adopted for mobile robots in~\cite{RosenfelderEbelEberhard21}.
At every time step $t\leq N_\textnormal{end}$ where $N_\textnormal{end}$ is the number of time steps considered in the experiment, each robot conducts a line search to find a candidate input sequence optimizing only its own input, then exchanges this prediction with its neighboring systems and finally ensures the solution is convex-like for each iteration $p \leq \bar p$, where $\bar p$ is the number of iterations per time step.
Due to the fixed iteration limit, solutions may be suboptimal.
In the studied setup, each robot is each other robot's neighbor, resulting in a maximally challenging communication topology.
At $p = \bar p$, each robot applies the first input of the found candidate input sequence.
The techniques explored in this work reduce the inter-agent communication by encoding the candidate input sequence prior to publication and each encoded sequence is decoded once received.

\section{Neural Network Architectures for Communication Reduction}\label{sec:nnarchitectures}

\noindent To address the communication challenges in DMPC, in the following, five different neural-network architectures are used to reduce the inter-agent communication in the distributed control algorithm from~\cite{StewartWrightRawlings11}.
An overview over the architectures, which all have an LSTM-based encoder in common, is given in~\cref{tab:architectureoverview}.
\begin{table*}[btp]
    \centering
        \caption{Overview over the considered models providing their name, encoder and decoder architectures as well as the quantity they communicate}
    \label{tab:architectureoverview}
  	\footnotesize
    \begin{tabular}{c c c c}\toprule
         Name & Encoder & Decoder & Communicate \\\midrule
         SHRED~\cite{WilliamsZahnKutz24} & LSTM & FFL  & network output \\
         Cell LSTM & LSTM & LSTM & cell state  \\
         LSTMP~\cite{SakSeniorBeaufays14} & LSTM projected & LSTM projected & network output and cell state   \\
         LSTM+FFL & LSTM and FFL & LSTM and FFL &  network output and cell state  \\
         LSTM+FFLH & LSTM & LSTM and FFL & network output \\\bottomrule
    \end{tabular}
\end{table*}
Other than the shallow recurrent decoder (SHRED)~\cite{WilliamsZahnKutz24}, which uses a standard feedforward neural network (FNN) as decoder, the networks use an LSTM not only in the encoder but also as part of the decoder.
SHRED performed well in~\cite{WilliamsZahnKutz24} for the reconstruction of high-dimensional states from senor data motivating an attempt to use it in a novel application.

Using an encoder-decoder LSTM~\cite{ChoEtAl14} architecture allows models to be trained and employed for multiple horizon lengths using a single training run.
The use of different encoder-decoder architectures is motivated by limited time available for the encoding and decoding operations as well as restricted data transfer from encoder to decoder.
The former limits the layer sizes and depth of the network, whereas the latter confines the amount of information from the encoder the decoder has available.
It is uncertain whether communicating the cell state, the output, or a combination of the two can capture the most amount of information in the latent space.
Thus, networks differ slightly depending on the quantity representing the reduced representation.
Communicating a combination of cell state and output leads to the introduction of projection layers to reduce the dimension of the hidden state and network output in some networks.
In~\cref{tab:architectureoverview}, these layers are indicated by projected and FFL (feed-forward layer), respectively.
The autoencoder from~\cite{SchizEbel25} serves as a baseline for reduced communication.

\begin{figure*}[btp]
	\centering
    \includegraphics{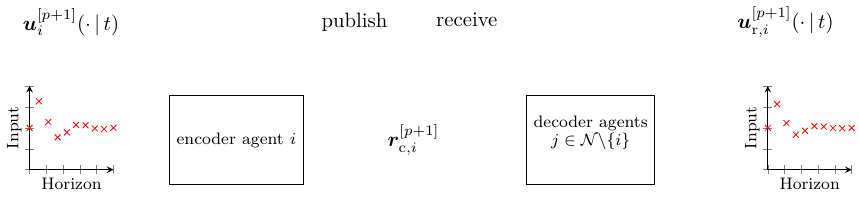}
    \caption{Semantic encoder-decoder structure: The encoded input sequence is published by robot $i$ and received and decoded by the other robots in the formation $j\in \mathcal N\backslash\lbrace i\rbrace$.}
    \label{fig:architecture_overview}
\end{figure*}

\begin{figure}[t]
    \centering
    \scalebox{0.7}{
     \includegraphics{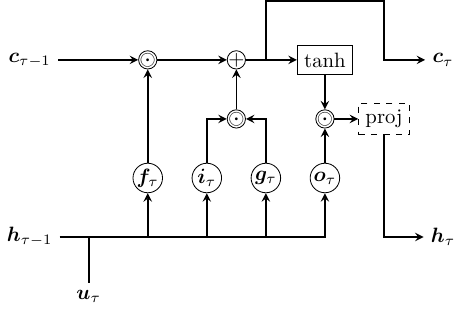}
     }
    \caption{Sketch of an encoder LSTM cell with optional projection layer}
    \label{fig:cellencoder}
\end{figure}

\begin{figure}[t]
\subfloat[\label{fig:SHRED_CELL}]{
    \begin{minipage}[t]{0.2\textwidth}
    \centering
    \scalebox{0.7}{%
     \includegraphics{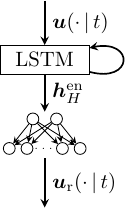}
         }
     \end{minipage}
}\hfil
\subfloat[\label{fig:lstmcell_CELL}]{
    \begin{minipage}[t]{0.2\textwidth}
    \centering
    \scalebox{0.7}{%
     \includegraphics{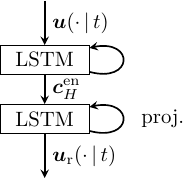}
        }    
     \end{minipage}
 }\\
 \subfloat[\label{fig:lstmproj_CELL}]{
\begin{minipage}[t]{0.2\textwidth}
    \centering
    \scalebox{0.7}{%
     \includegraphics{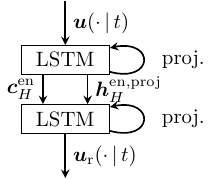}
         }
     \end{minipage}
     }\hfil
 \subfloat[\label{fig:LSTMfnn_cell}] {
    \begin{minipage}[t]{0.2\textwidth}
    \centering
    \scalebox{0.7}{%
     \includegraphics{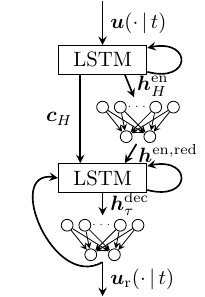}
         }
         \end{minipage}
}
\caption{Overview over different LSTM architectures: (a)~SHRED, (b)~Cell LSTM, (c)~LSTMP, (d)~LSTM+FFL}\label{fig:archover}
\end{figure}

To reduce the inter-agent communication, the predicted input sequence along the prediction horizon $\bm u_{i}^{[p+1]}(\cdot \vert t)$ at iteration~$p$ and time step~$t$  of each agent $i\in \mathcal N$ is compressed into a latent space using a pre-trained encoder prior to the publication of the message.
The reduced representation $\bm r_{\mathrm{c},i}^{[p+1]}(\cdot \vert t)$ is communicated and when another robot receives the message, the predicted input sequence is reconstructed using a decoder, as shown in \cref{fig:architecture_overview}.
The same pre-trained model is used in all robots and each message is reduced and reconstructed individually, making the communication reduction independent of the formation size.
Given the LSTM-input $\bm u_\tau$, one forward pass through an LSTM cell at index $\tau$ corresponds to
\begin{align}
    \bm i_\tau &= \sigma(\bm W_{\mathrm{ii}} \bm  u_\tau + \bm  b_{\mathrm{ii}} + \bm W_{\mathrm{hi}} \bm h_{\tau-1} + \bm b_{\mathrm{hi}}) \\
    \bm f_\tau &= \sigma(\bm W_\mathrm{if} \bm  u_\tau + \bm b_\mathrm{if} + \bm W_\mathrm{hf} \bm h_{\tau-1} + \bm b_\mathrm{hf}) \\
    \bm g_\tau &= \tanh(\bm W_\mathrm{ig} \bm u_\tau + \bm b_\mathrm{ig} + \bm W_\mathrm{hg} \bm h_{\tau-1} + \bm b_\mathrm{hg}) \\
    \bm o_\tau &= \sigma(\bm W_\mathrm{io} \bm u_\tau + \bm b_\mathrm{io} + \bm W_\mathrm{ho} \bm h_{\tau-1} + \bm b_\mathrm{ho}) \\
    \bm c_\tau &= \bm f_\tau \odot \bm c_{\tau-1} + \bm i_\tau \odot \bm g_\tau \\
    \bm h_\tau &= \bm o_\tau \odot \tanh(\bm c_\tau)
\end{align}
with the input gate $\bm i_\tau$, the forget gate $\bm f_\tau$, the cell gate $\bm g_\tau$ and the output gate $\bm o_\tau$.
Further, $\bm c_\tau$ corresponds to the cell state, $\bm h_\tau$ to the hidden state, $\sigma(x) = 1 / (1+ \exp(-x))$ is the sigmoid function, and $\odot$ the Hadamard product.
The quantities $\bm W_{lm}$ and $\bm b_{lm}$ for $l\in\lbrace\mathrm{i},\mathrm{h}\rbrace$ and $m\in\lbrace\mathrm{i},\mathrm{f},\mathrm{g},\mathrm{o}\rbrace$ are the weights and biases, respectively.
This basic LSTM architecture is modified slightly, depending on the neural network architecture used, e.g., by adding a projection layer.
In the forward pass, for each robot $i$ at time $t$ and iteration $p$ it holds that $\bm u_{\tau,i}^\mathrm{en} = \bm u_{i}^{[p+1]}(t+k\,\vert\,t)$ for $(\tau,k)\in\lbrace(1,0),(2,1),\dots,(H,H-1)\rbrace$.
A cell of this LSTM encoder is displayed in \cref{fig:cellencoder}.
As the same network is used for each agent, the robot index $i$ is omitted in the following for variables related to neural networks.
SHRED uses a standard LSTM encoder combined with a feed-forward decoder.
The use of an FNN as decoder requires retraining when the length of the prediction horizon is changed.
The output of the encoder, i.e., $\bm r_\mathrm{c} = \bm h_H$, is communicated and used as input by the decoder.
A sketch of this architecture is presented in \cref{fig:SHRED_CELL}.

In the variants using LSTMs also in the decoder, the input predictions are reconstructed in reverse order, such that $\bm u_{\mathrm{r}}(t+H-1-l\,\vert\,t) = \bm u_\tau^\mathrm{de} $ for $(\tau,l)\in\lbrace(1,0),(2,1),\dots,(H,H-1)\rbrace$.
The first input $\bm u_{\tau=0}^\mathrm{de}$ is not used in the reconstruction of the control input vector but only used as an LSTM input.

The variant denoted as cell LSTM communicates only the cell state of the encoder $\bm r_{\mathrm{c}} = \bm c_H^\mathrm{en}$.
The decoder sets the initial input and hidden state to zero, i.e., $\bm h_0^\mathrm{de} = \bm 0$ and $\bm u_0^\mathrm{de} = \bm 0$, and uses the cell state it received as the initial cell state meaning that $\bm c_H^\mathrm{de} = \bm r_{\mathrm{c}}$.
An overview of this architecture is given in \cref{fig:lstmcell_CELL}.

The LSTMP architecture, based on LSTM projected~\cite{SakSeniorBeaufays14} and depicted in \cref{fig:lstmproj_CELL}, communicates the cell state and projected hidden state, i.e., $\bm r_\mathrm{c} = \begin{bmatrix}
    \bm c_{H}^\mathrm{en}{}\tran & \bm h_{\mathrm{proj}, {H}}^\mathrm{en}{}\tran
\end{bmatrix}\tran$.
In this model, the output of the LSTM is projected in both the encoder and the decoder to reduce its dimension to the dimension of a velocity pair using a projection layer of the form $\bm h_{\mathrm{proj},\tau} = \bm W_\mathrm{proj}\bm h_\tau$ where $\bm W_\mathrm{proj}$ is a learnable weight matrix.

The projected hidden state is used as the input to the decoder, i.e., $\bm u_0^\mathrm{de} = \bm h_{\mathrm{proj}, {H}}^\mathrm{en}$, and the last cell state of the encoder is used as the initial first cell state of the decoder meaning that $\bm c_{0}^\mathrm{de}=\bm c_{H}^\mathrm{en}$.
The initial hidden states are set to zero.

\begin{figure}[t]
    \centering
    \scalebox{0.7}{%
     \includegraphics{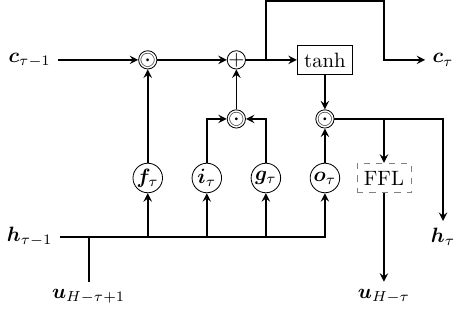}
         }
    \caption{Sketch of a decoder LSTM with an FFL. The FFL is only applied to the output of the last layer at $\tau = H$. Note that the prediction sequence is recreated in reverse order from $H$ to $0$ in the decoder.}
    \label{fig:cellfnndecoder}
\end{figure}

In the LSTM+FFL variant, sketched in \cref{fig:LSTMfnn_cell}, instead of using a projection in each pass of the LSTM, only the output of the encoder is projected to a two-dimensional representation using a single-layer FNN, i.e.,
\begin{align}
    \bm h_{\mathrm{FFL}, H} = \bm W_\mathrm{FFL} a({\bm h_{H}}) + \bm b_\mathrm{FFL}
\end{align}
with the weight matrix $\bm W_\mathrm{FFL}$, the bias vector $\bm b_\mathrm{FFL}$, and the activation function $a$.
This modification of LSTMP allows to use an unprojected hidden state while still communicating an output projected to the size of a velocity pair.
The projected output and the cell state are communicated so that $\bm r_\mathrm{c} = \begin{bmatrix}
    \bm c_{H}^\mathrm{en}{}\tran & \bm h_{\textrm{FFL}, H}^\mathrm{en}{}\tran
\end{bmatrix}\tran$.
A decoder layer of this architecture is presented in \cref{fig:cellfnndecoder}.
The decoder uses the communicated cell state as the initial cell state and the projected output as the initial input while the initial hidden states are set to zero.

A similar architecture has also been implemented to communicate only the output, i.e., $\bm r_\mathrm{c} = \bm h_H^\mathrm{en}$, in the following called LSTM+FFLH.
The received output is used as the initial hidden state and additionally it is projected using an FFL to serve as the initial input.
To act as the input of the next pass, the output of a forward pass is reduced using the same single-layer FNN as is used to reduce the received output.
The cell state is initially set to zero.
Note that, in multi-layer LSTMs, only the cell or network output of the last layer is communicated and consequently also only the respective states of the first layer in the decoder are set to a non-zero vector.

\section{Data Generation, Tuning, and Training}\label{sec:datageneration}
\noindent Three main steps are taken to employ a learning-based communication strategy for DMPC, namely data generation by recording inter-agent communication, training and tuning of the encoder-decoder architecture as a standalone model, and, finally, splitting the model into its components and implementing the encoder prior to the transmission and the decoder post reception.
The dataset consists of the recorded candidate control inputs $\bm u^{[p]}(\cdot \,\vert\,t)$ for $p\in\mathbb Z_{0:\bar p}$ for each $t\in \mathbb Z_{0:N_\textnormal{end}}$ published by a single agent with the number of time steps~$N_\textnormal{end}\coloneqq \lceil \Delta t^{-1} T_\textnormal{end}\rceil$ where $T_\textnormal{end}$ is the simulation duration.
During training, the control logic of each robot runs in a separate process but all processes run on a shared compute unit.
For the data generation, only formations consisting of two robots have been used.
Using only the communicated predictions of one robot from two-robot formations is sufficient as in simulations without disturbances, all robots in a formation make similar predictions and cover the input space similarly well.
The initial pose of the geometric center is uniformly randomly chosen from $x,y\in[\qty{-10}{\meter},\qty{10}{\meter})$ and $\theta\in[\SI[parse-numbers=false]{-\pi}{\radian},\SI[parse-numbers=false]{\pi}{\radian})$.
The robots are placed with $x_\mathrm{rel} = \pm \qty{0.5}{\meter}$ relative to the geometric center using the same $y$-position and orientation as the geometric center.
As, for control, only the relative positioning (i.e., the error) toward the goal formation-pose matters, it is enough to vary either the starting or the goal poses, where we opt for the former.
The desired geometric center is always set to $\hat{\bm y}_\mathrm{d} = \begin{bmatrix}\qty{0}{\meter} & \qty{0}{\meter} & \qty{0}{\radian}\end{bmatrix}\tran$.

For horizon length $H=20$, $2000$ randomly generated scenarios are used in the training set.
Data generated from these scenarios is used to train the neural network architectures with fixed length.
To train encoder-decoder LSTMs for variable horizon lengths, data is additionally recorded for $200$ scenarios for each horizon length $H\in\mathbb Z_{21:25}$.
The $200$ scenarios are a random subset of the $2000$ scenarios used for $H=20$.

The simulation duration for the data generation is set to $T_\textnormal{end}= \qty{50}{\second}$. This gives the formations enough time to approach the setpoint and ensures that robots do not spend too much time in close proximity to the desired pose.
For the data generation, training, and when testing the performance of the communication reduction, the following setup is used if not stated otherwise.
The distributed controller is implemented in C++ using CasADi~\cite{AnderssonEtAl19} to perform automatic differentiation and Eigen~\cite{GuennebaudJacob10} for linear algebra.
Robots communicate via UDP multicast using ZCM~\cite{ZCMContributors21}, which is based on LCM~\cite{HuangOlsonMoore10}.
The initial step size $\bar \alpha = 0.1$, the backtracking factor $\beta = 0.5$, and the Armijo factor $\sigma = 0.5$ are used based on previous experience with the employed distributed optimization algorithm~\cite{RosenfelderEbelEberhard22}.
The number of iterations per time step is set to $\bar p = 3$ and the sampling time is $\Delta t = \qty{0.25}{\second}$.
The first candidate inputs of the optimization at the end of each time step, i.e., $\bm u^{[\bar p]} (t\,\vert\, t)$, are communicated to a simulator program written in Python, which updates the robots' poses and sends the updated poses back.
The simulator uses a multi-body model, previously used in~\cite{EbelEtAl22, EbelRosenfelderEberhard24, SchizEbel25} including actuation dynamics and inertia of the chassis and the wheels, and hence having been found to translate well to real-world experiments.
This introduces a model mismatch between prediction and simulation of the robot motions.

For each neural network architecture, hyperparameter tuning is performed.
The models are implemented in PyTorch~\cite{AnselEtAl24} and training runs are tracked using Weights \& Biases~\cite{Biewald20}.
The dataset is split into $90\%$ training data and $10\%$ validation data.
Employing the encoder-decoder structure in the optimization algorithm and running formation tasks is considered the test phase, see \cref{sec:simulativeresults}.
The initial learning rate is set to either $0.001$ or $0.01$ combined with a reduce-on-plateau learning rate scheduler, reducing the learning rate by a factor of $0.1$ if the validation loss does not decrease for five epochs.
With this approach, the learning has mostly ceased after $200$ epochs, such that $200$ is chosen as an upper epoch limit for all tuning runs.
For training, the batch sizes 265, 512, 1024 are considered.
The training uses the ADAM optimizer~\cite{KingmaBa14} and the mean-square error loss function.
In all neural network model setups, it is ensured that the code size, i.e., the dimension of the latent space, is 10 corresponding to the code size used for the autoencoder in the prior work~\cite{SchizEbel25}, to have an ample baseline.
This means that, in cases where only one of the states is communicated, the hidden and cell state size is 10.
This is the case for SHRED, cell LSTM, and LSTM+FFLH.
In other cases, where a cell state and the network output are communicated together, the hidden and cell state size is set to eight and the output size is two, as it represents an extended term of the prediction.
This applies to LSTMP and LSTM+FFL.
LSTMs are considered with one to three layers.
For the LSTM+FFL architecture, a single FNN layer is considered with input size eight and output size two.
As activation functions, hyperbolic tangent and ReLU are considered.
The FNN layer in the LSTM+FFLH has an input size of 10 and an output size of two.
The FNN decoder of SHRED is tuned using the hyperbolic tangent, ReLU, and leaky ReLU activation functions.
The decoder has two hidden layers.
The inner layer, i.e., the hidden layer closer to the code, has one of the sizes $25$, $23$, $20$, $18$, or $15$.
The outer layer,  i.e., the hidden layer closer to the output, is either $35$, $33$, $30$, $28$, or $26$.
The small layer sizes are chosen to ensure a low computational overhead.

The best parameter setup found for each variant are summarized in \cref{tab:besthyperparameter}.
\begin{table*}[btp]
    \centering
    \caption{Hyperparameter settings with lowest mean validation loss over an epoch for the considered architectures.}
    \label{tab:besthyperparameter}
    \footnotesize
    \begin{tabular}{c c c c c c}\toprule
         & SHRED & Cell LSTM & LSTMP & LSTM+FFL & LSTM+FFLH \\\midrule
         LSTM layer & $2$ & $2$ & $2$ & $2$ & $2$   \\
         Activation function & tanh &  & & tanh/relu &  tanh  \\
         Inner layer & $25$ & & & &   \\
         Outer layer & $35$ & & & &   \\
         batch size & $1024$ & $256$ & $256$ & $256$ & $256$ \\
         Validation Loss & $0.00039$ & $0.00093$ & $0.00130$ & $0.000574$/$0.000568$ & $0.00059$ \\\bottomrule
    \end{tabular}
    
\end{table*}
For the LSTM+FFL variant, the ReLU activation function worked slightly better in training but using the hyperbolic tangent outperformed the ReLU activation function in testing.
The validation loss row shows that the SHRED-based variant yielded the best whereas the LSTMP variant yielded the worst training performance.
When considering only architectures trained for a variety of prediction horizons, the LSTM+FFL combination seems to train best.

\section{Test: Simulative Analysis}\label{sec:simulativeresults}
\noindent This section focuses on testing the convergence performance of the optimization algorithm when incorporating the trained models and reducing the communication under idealized communication conditions.
This corresponds to the test phase of the model training.
For the analysis in this section, all processes run on a single computer and no communication between physical devices takes place.
No packets are lost in this idealized setup.
Afterwards, in the subsequent section, the focus shifts to determining how different communication approaches perform under demanding and realistic wireless communication conditions.

\begin{figure}[t]
    \centering
\includegraphics{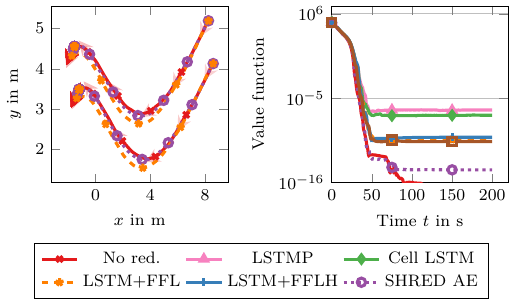}
    \caption{Left: Trajectories of an example scenario with two robots for $H=20$ using no reduction (No red.) in the communication, SHRED, and LSTM+FFL communication reduction. The arrows in the left plot indicate the robot orientations for the task with full communication with the opacity increasing with time. Right: The corresponding value functions including all tested communication architectures for one robot.}
    \label{fig:exemplarytask}
\end{figure}

\begin{figure}[t]
    \centering
    \includegraphics{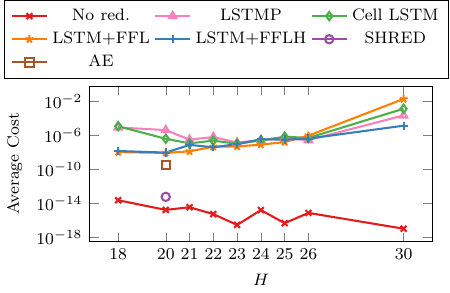}
    \caption{Mean cost values at $T_\textnormal{end}$ of one robot for the test scenarios for all communication reduction techniques. The horizon lengths are connected with lines to improve readability.}
    \label{fig:finalaveragecostfordiffH}
\end{figure}

To quantify the performance, $200$ scenarios for formation sizes from two to six robots are considered for $H=20$.
The simulation duration $T_\textnormal{end}=\qty{200}{\second}$ is used for formations of two to four robots while larger formations require more time to allow testing convergence empirically. Thus, a simulation duration of $T_\textnormal{end} = \qty{300}{\second}$ and $T_\textnormal{end} = \qty{450}{\second}$ is used for formations of five and six robots respectively.
Initial and desired pose of the geometric center are chosen randomly with uniform distribution for each scenario within the regions
$x,y\in [\qty{-20}{\meter},\qty{20}{\meter})$ for the position and $\theta\in[\SI[parse-numbers=false]{-\pi}{\radian},\SI[parse-numbers=false]{\pi}{\radian})$ for the orientation.
A test scenario consisting of a formation with two robots is exemplarily displayed in \cref{fig:exemplarytask}.
As an additional baseline, the retrained autoencoder from~\cite{SchizEbel25} is included.
The summary of the average performances of the different communication techniques is presented in \cref{fig:finalaveragecostfordiffH}.
For $H=20$, the horizon length contributing the most samples to the training dataset, SHRED outperforms the other techniques.
Of the variants compatible with and trained for a variable horizon length, the combination of LSTM and FNN performed better than cell LSTM and LSTMP.
This matches the training results.
Looking at the best and worst performance confirms this observation, see \cref{tab:finalcoststatisticsH20}.
\begin{table}[tbp]
    \centering
    \caption{Overview of cost at $T_\textnormal{end}$ for $H=20$.}
    \label{tab:finalcoststatisticsH20}
    \footnotesize
    \begin{tabular}{l c c c }\toprule
         & Avg. & Max. & Min.  \\\midrule
        No reduction &  $1.66\cdot10^{-15}$ & $9.38\cdot10^{-14}$ & $< 10^{-16}$\\ 
        AE & $3.12\cdot 10^{-10}$ & $9.89\cdot 10^{-9}$ &  $2.03\cdot 10^{-11}$ \\
        SHRED & $5.76\cdot 10^{-14}$ & $2.82\cdot 10^{-13}$ & $4.20\cdot 10^{-15}$  \\
        Cell LSTM & $4.14\cdot 10^{-7}$ & $7.29\cdot 10^{-6}$ & $5.33\cdot 10^{-8}$  \\
        LSTMP & $4.41\cdot 10^{-6}$ & $2.13\cdot 10^{-5}$ & $2.69\cdot 10^{-7}$  \\
        LSTM+FFL & $8.89\cdot 10^{-9}$ & $4.90\cdot 10^{-7}$ & $2.99\cdot 10^{-11}$  \\
        LSTM+FFLH & $9.66\cdot 10^{-9}$ & $6.74\cdot 10^{-7}$ & $6.01\cdot 10^{-11}$  \\\bottomrule
    \end{tabular}
\end{table}
That methods trained for variable horizon lengths perform worse on a specific horizon than methods specifically trained for this horizon length is expected as they sacrifice some accuracy for generality.
The findings for $H=20$ are further supported by the value function plot on the right of~\cref{fig:exemplarytask} for one example scenario.
The value function corresponds to the cost function values at the last iterate of each time step, i.e., the (sub-)optimal solution.
\begin{figure}[btp]
    \centering
    \includegraphics{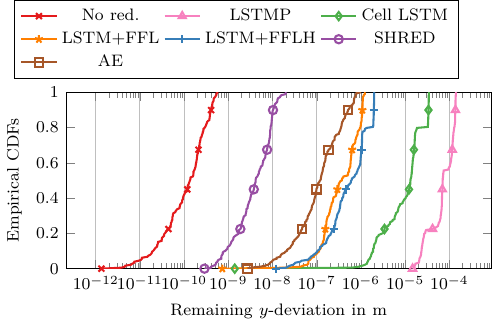}
    \caption{Empirical cumulative distribution functions~(CDFs) of error in $y$-direction for $60$ random formation tasks with full and reduced communication.}
    \label{fig:cfd_y_direction}
\end{figure}
An important performance measure specifically for the considered nonholonomic mobile robots is the error in the hard-to-control lateral-direction, which is the $y$ direction at the setpoint.
To measure the performance when using different communication techniques, the cumulative density functions~(CDFs) of the $y$-error at $T_\textnormal{end}=\qty{300}{\second}$ of $60$ different scenarios consisting of formations of two to four robots are compared.
The CDFs are depicted in \cref{fig:cfd_y_direction}.
The formations depicted in the CDF plot cover a smaller area than the scenarios used for the cost comparison.
The CDFs confirm the previous observations, full communication achieves the lowest $y$-error with $100\%$ reaching a deviation of less than $1.0 \cdot 10^{-9}$~\si{\meter}, closely followed by SHRED.
SHRED's $y$-error is almost two orders of magnitude lower than the autoencoder error.
The combinations of LSTM and FNN come close to the performance of the autoencoder towards $100\%$ while the other LSTM variants barely reach an accuracy below \qty{1}{\mm}.

For each of the horizon lengths $H\in\lbrace 18, 21, 22, 23, 24, 25, 26, 30\rbrace$, the same tasks are used for testing.
Looking at the average cost at $T_\textnormal{end}$ for different prediction horizons included in the training scenarios in \cref{tab:finalcosth1830}, the architectures combining an LSTM element with an FFL projection have the lowest average cost at $T_\textnormal{end}$ in most of the horizon length covered by the training.
The same behavior can be observed for $H=18$, a horizon length outside of the training range.
However, for increasingly longer horizons, the cell LSTM and LSTMP variant begin to have lower cost values at $T_\textnormal{end}$ compared to the LSTM+FFL variant.
For $H=30$, well outside the prediction horizon training range, the LSTM+FFL variant performs worst, but the LSTM+FFLH variant retains best performance.
The CDFs for the $y$-deviation under equal conditions as used for $H=20$ show that LSTMP and cell LSTM perform better for longer prediction horizons, even slightly outside the training range.
At $H=26$, all architectures experience similar convergence behavior.
The LSTM+FFLH model performs similarly for all tested horizon lengths with $100\%$ of scenarios reaching a $y$-deviation of about $10^{-5}$~\si{\meter}, even at $H=30$ where the other models perform significantly worse and where the LSTM+FFL model does not achieve an accuracy smaller than \qty{1}{\mm} for any of the considered scenarios.
\begin{figure*}[btp]
    \centering
\includegraphics{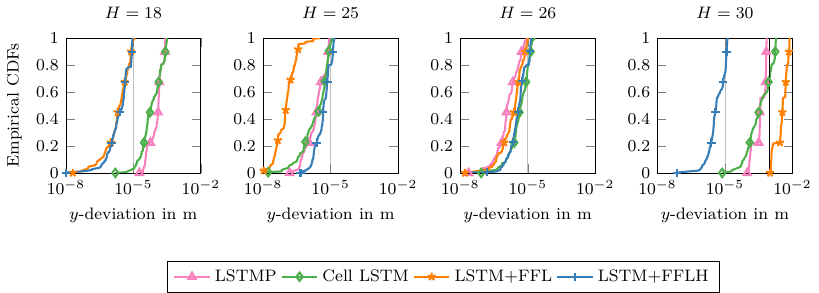}
\caption{Empirical cumulative distribution functions~(CDFs) of error in $y$-direction for \num{60} random formation tasks with encoder-decoder LSTM-reduced communication for different $H$ values.}\label{fig:ydevhdiff}
\end{figure*}

\begin{table}[t]
    \centering
    \caption{Mean, max, and min cost values at $T_\textnormal{end}$ of one robot for the test scenarios for all communication reduction techniques working for a variable horizon length. For each horizon length, the best performing reduction method is highlighted.}
    \label{tab:finalcosth1830}
    \footnotesize
    \begin{tabular}{l c c c c}\toprule
         & H & Avg. & Max & Min \\\midrule
        No reduction & 18 & $2.29\cdot 10^{-14}$ & $2.26\cdot 10^{-12}$ & $<10^{-16}$ \\
        Cell LSTM & 18 & $1.24\cdot 10^{-5}$ & $5.25\cdot 10^{-5}$ & $1.57\cdot 10^{-6}$ \\
        LSTMP & 18 & $8.92\cdot 10^{-6}$ &  $3.054\cdot 10^{-5}$ &  $6.45\cdot 10^{-7}$ \\        
        LSTM+FFL & 18 & \cellcolor{gray!30} $1.11\cdot 10^{-8}$ & \cellcolor{gray!30}$6.55\cdot 10^{-8}$ & \cellcolor{gray!30}$4.28\cdot 10^{-10}$ \\
        LSTM+FFLH & 18 & $1.50\cdot 10^{-8}$ & $2.34\cdot 10^{-7}$ & $9.81\cdot 10^{-10}$ \\ \addlinespace[0.3em]
        No reduction & 21 & $3.42\cdot 10^{-15}$ & $3.08\cdot 10^{-13}$ & $<10^{-16}$ \\
        Cell LSTM & 21 & $1.20\cdot 10^{-7}$ & $2.26\cdot 10^{-6}$ & $1.09\cdot 10^{-8}$ \\ 
        LSTMP & 21 & $3.20\cdot 10^{-7}$ &  $5.11\cdot 10^{-6}$ &  $2.99\cdot 10^{-8}$ \\
        LSTM+FFL & 21 & \cellcolor{gray!30}$1.36\cdot 10^{-8}$ & \cellcolor{gray!30}$7.06\cdot 10^{-7}$ & $8.22\cdot 10^{-11}$ \\
        LSTM+FFLH & 21 & $7.84\cdot 10^{-8}$ & $5.89\cdot 10^{-6}$ & \cellcolor{gray!30}$7.09\cdot 10^{-11}$ \\ \addlinespace[0.3em]
        No reduction & 22 & $5.27\cdot 10^{-16}$ & $3.02\cdot 10^{-14}$ & $<10^{-16}$ \\
        Cell LSTM & 22 & $2.48\cdot 10^{-7}$ & $1.08\cdot 10^{-5}$ & $1.40\cdot 10^{-8}$ \\
        LSTMP & 22 & $6.68\cdot 10^{-7}$ &  $3.59\cdot 10^{-6}$ &  $3.69\cdot 10^{-8}$ \\
        LSTM+FFL & 22 & $4.82\cdot 10^{-8}$ & $3.11\cdot 10^{-6}$ & $7.39\cdot 10^{-11}$ \\
        LSTM+FFLH & 22 & \cellcolor{gray!30}$4.13\cdot 10^{-8}$ & \cellcolor{gray!30}$2.08\cdot 10^{-6}$ & \cellcolor{gray!30}$4.62\cdot 10^{-11}$ \\\addlinespace[0.3em]
        No reduction & 23 & $< 10^{-16}$ & $1.50\cdot 10^{-15}$ & $<10^{-16}$ \\
        Cell LSTM & 23 & $1.08\cdot 10^{-7}$ & $4.84\cdot 10^{-6}$ & $2.22\cdot 10^{-9}$ \\
        LSTMP & 23 & $1.47\cdot 10^{-7}$ &  $2.77\cdot 10^{-6}$ &  $1.76\cdot 10^{-8}$ \\
        LSTM+FFL & 23 & \cellcolor{gray!30}$5.10\cdot 10^{-8}$ & \cellcolor{gray!30}$2.57\cdot 10^{-6}$ & \cellcolor{gray!30}$5.44\cdot 10^{-11}$ \\
        LSTM+FFLH & 23 & $9.66\cdot 10^{-8}$ & $4.88\cdot 10^{-6}$ & $2.37\cdot 10^{-10}$ \\\addlinespace[0.3em]
        No reduction & 24 & $1.57\cdot 10^{-15}$ & $1.24\cdot 10^{-13}$ & $<10^{-16}$ \\
        Cell LSTM & 24 & $2.72\cdot 10^{-7}$ & $1.29\cdot 10^{-5}$ & $3.94\cdot 10^{-9}$ \\
        LSTMP & 24 & $2.83\cdot 10^{-7}$ &  $1.52\cdot 10^{-5}$ &  $1.01\cdot 10^{-8}$ \\
        LSTM+FFL & 24 & \cellcolor{gray!30}$8.43\cdot 10^{-8}$ & \cellcolor{gray!30}$5.59\cdot 10^{-6}$ & \cellcolor{gray!30}$2.24\cdot 10^{-10}$ \\
        LSTM+FFLH & 24 & $3.61\cdot 10^{-7}$ & $2.72\cdot 10^{-5}$ & $1.40\cdot 10^{-9}$ \\\addlinespace[0.3em]
        No reduction & 25 & $< 10^{-16}$ & $3.10\cdot 10^{-15}$ & $<10^{-16}$ \\
        Cell LSTM & 25 & $7.16\cdot 10^{-7}$ & $6.14\cdot 10^{-5}$ & $5.27\cdot 10^{-9}$ \\
        LSTMP & 25 & $4.61\cdot 10^{-7}$ &  $3.53\cdot 10^{-5}$ &  $9.75\cdot 10^{-9}$ \\
        LSTM+FFL & 25 & \cellcolor{gray!30}$1.69\cdot 10^{-7}$ & \cellcolor{gray!30}$9.95\cdot 10^{-6}$ & \cellcolor{gray!30}$3.16\cdot 10^{-12}$ \\
        LSTM+FFLH & 25 & $3.30\cdot 10^{-7}$ & $2.13\cdot 10^{-5}$ & $3.05\cdot 10^{-9}$ \\\addlinespace[0.3em]
        No reduction & 26 & $7.30\cdot 10^{-16}$ & $5.05 \cdot 10^{-14}$ & $<10^{-16}$ \\
        Cell LSTM & 26 & $5.31\cdot 10^{-7}$ & $3.07\cdot 10^{-5}$ & $2.09\cdot 10^{-8}$ \\ 
        LSTMP & 26 & \cellcolor{gray!30}$2.86\cdot 10^{-7}$ &  $2.01\cdot 10^{-5}$ &  $9.02\cdot 10^{-9}$ \\
        LSTM+FFL & 26 & $9.66\cdot 10^{-7}$ & \cellcolor{gray!30}$1.50\cdot 10^{-5}$ & $1.90\cdot 10^{-7}$ \\
        LSTM+FFLH & 26 & $4.39\cdot 10^{-7}$ & $3.13\cdot 10^{-5}$ & \cellcolor{gray!30}$2.55\cdot 10^{-9}$ \\\addlinespace[0.3em]
        No reduction & 30 & $<10^{-16}$ & $6\cdot 10^{-16}$ & $<10^{-16}$ \\
        Cell LSTM & 30 & $0.00146$ & $0.00457$ & $0.00038$ \\
        LSTMP & 30 & $0.000238$ &  $0.00066$ &  $3.40\cdot 10^{-5}$ \\
        LSTM+FFL & 30 & $0.02069$ & $0.05110$ & $0.00380$ \\
        LSTM+FFLH & 30 & \cellcolor{gray!30}$1.45\cdot 10^{-5}$ & \cellcolor{gray!30}$0.00014$ & \cellcolor{gray!30}$2.92\cdot 10^{-6}$ \\\bottomrule
    \end{tabular}
\end{table}

\section{Numerical Experiments on Embedded Hardware}\label{sec:numexperiments}
\begin{figure}[btp]
    \centering
    \includegraphics[width=0.5\linewidth]{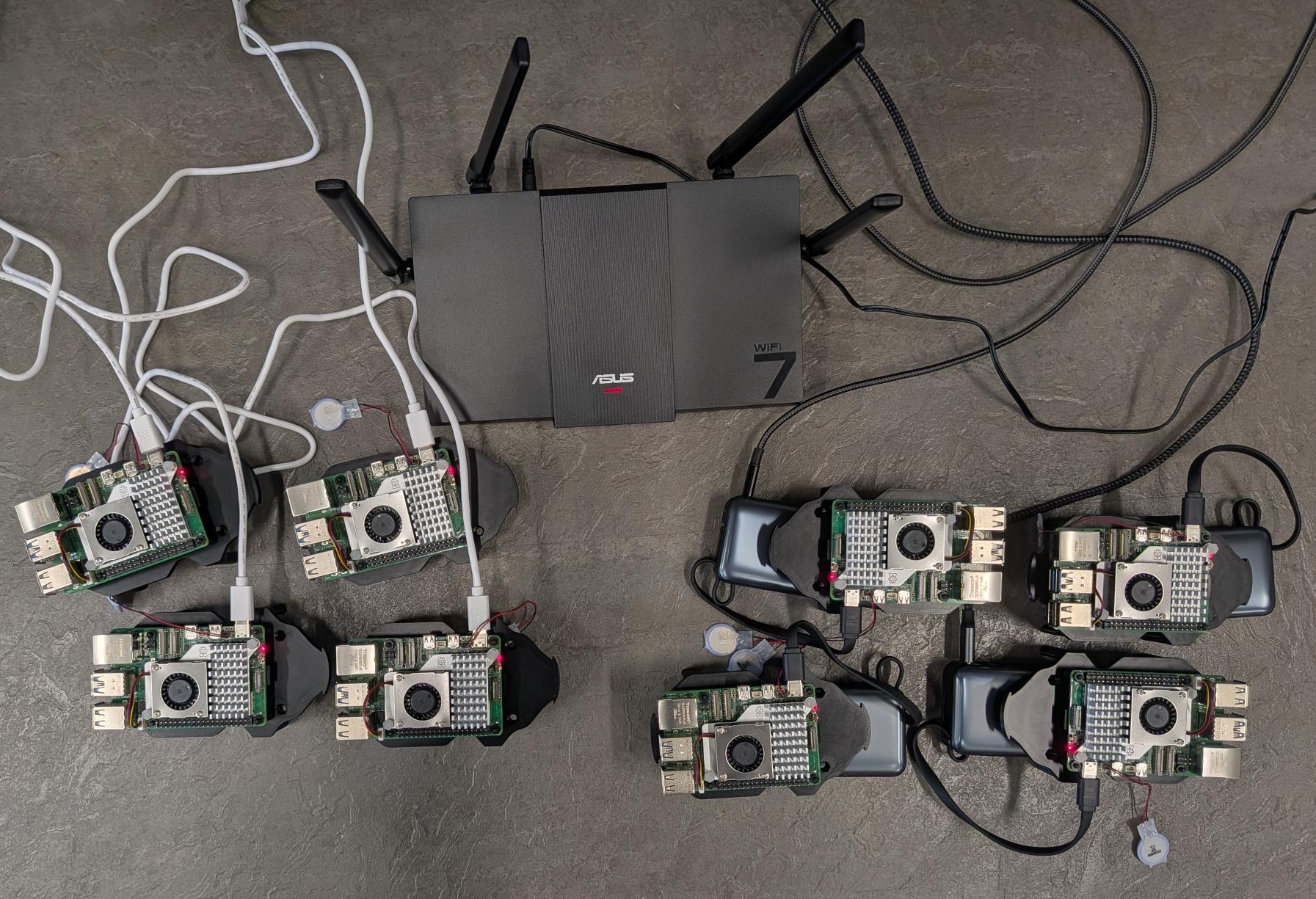}
    \caption{Setup for the numerical experiments}
    \label{fig:expsetup}
\end{figure}

\begin{figure}[btp]
    \centering
     \includegraphics{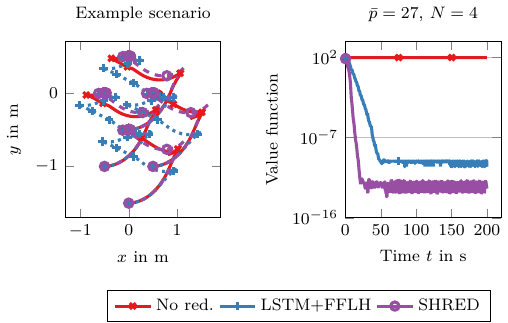}
    \caption{Left: Trajectories of an example parallel parking scenario with four robots using no reduction (No red.) in the communication, SHRED, and LSTM+FFLH communication reduction. Right: The value functions of one of the robots of a scenario with four robots.}
    \label{fig:embfour}
\end{figure}

\noindent To compare full communication to semantically reduced communication in a realistic communication setting, the distributed optimization algorithm is run on Raspberry Pi 5 single-board computers with 8 GB RAM, as they are also fitted to the robots depicted in~\cref{fig:robexample}.
The optimization algorithm for one robot is run on a dedicated Raspberry Pi.
The communication takes place via \qty{5}{\GHz} Wi-Fi using UDP multicast.
Each robot re-publishes messages once if it has not received all messages it is waiting for after half of the time dedicated to an iteration.
Additionally, robots use the best information available, i.e., if no new message is received, old information is reused and if information of a subsequent iteration but not of the current iteration is available, the information meant for the next iteration is already used.
The experimental setup is exemplarily depicted in~\cref{fig:expsetup}.
The numerical experiment only considers simulated mechanical dynamics. 
This simplifies the experiment in two ways as the challenges of moving communication endpoints are neglected and so are deviations between prediction and the behavior of the real robot beyond model mismatches.
The simulator program runs on a personal computer with an Intel Core Ultra 5 238V @ 2.1 GHz and 32 GB of memory, which is connected to the router via LAN.
It receives the applied inputs from each robot and computes the new poses and publishes the updated poses.
This introduces additional communication but moves the application of the control input from the on-board computer to an external computer thus slightly modifying the setup.
As reduction methods only SHRED and LSTM+FFLH are considered as the best performing methods from~\cref{sec:simulativeresults}.
The choice of LSTM+FFLH over LSTM+FFL is motivated by the better generalization to larger values of the prediction horizon lengths.
Parallel parking tasks as depicted exemplarily for four robots in~\cref{fig:embfour} are used in the numerical experiments.

Each experiment (except for the case with $\bar p = 3$, as under these conditions all variants work well) was repeated a few times to account for a varying experimental environment that could be impacting the performance.
In the following, the best performing repetition is depicted in value-function plots.

\begin{figure}[btp]
    \centering
     \includegraphics{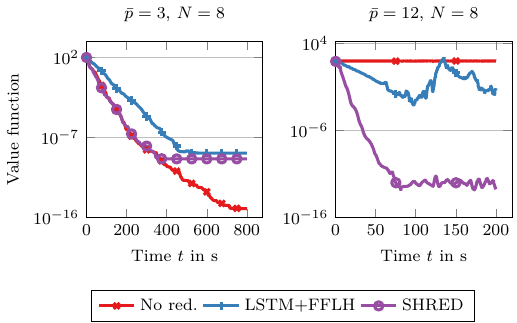}
    \caption{Value functions for two experiments with $N=8$ and $H=20$}
    \label{fig:embeddedexemplarytaskeight}
\end{figure}

In the first experiment, a formation of eight robots is chosen combined with $H=20$ and $\bar p=3$ aligning with the setup chosen for the simulative analysis.
Under these conditions, all three communication configurations manage to receive around $99\%$ of the amount of communicated information, see~\cref{tab:embeddedperformance}, and display convergence similar to the simulation results, as shown in~\cref{fig:embeddedexemplarytaskeight}.
\begin{table}[btp]
    \centering
    \caption{Average amount of communicated information received in percent on embedded hardware}
    \label{tab:embeddedperformance}
    \footnotesize
    \begin{tabular}{c c c c c c c}\toprule
         Name & $N$ & $T_\textnormal{end}$ & $H$ & $\bar p$ & Avg.\ info. received\ in $\%$ \\\midrule
          No red. & $8$ & $800$ & $20$ & $3$ & $98.97$  \\
          SHRED & $8$ & $800$ & $20$ & $3$ &  $99.02$  \\
          LSTM+FFLH & $8$ & $800$ & $20$ & $3$ & $99.12$ \\
          No red. & $4$ & $200$ & $20$ & $27$ & $0.00$   \\
          SHRED & $4$ & $200$ & $20$ & $27$ & $98.29$   \\
          LSTM+FFLH & $4$ & $200$ & $20$ & $27$ &  $90.70$ \\
          No red. & $8$ & $200$ & $20$ & $12$ & $0.02$  \\
          SHRED & $8$ & $200$ & $20$ & $12$ &  $98.01$  \\
          LSTM+FFLH & $8$ & $200$ & $20$ & $12$ & $87.43$ \\
          No red. & $8$ & $300$ & $25$ & $10$ & $31.41$ \\
          LSTM+FFLH & $8$ & $300$ & $25$ & $10$ & $93.38$ \\
          \bottomrule
    \end{tabular}
\end{table}%
Using a smaller formation of four robots and increasing the number of iterations to $\bar p=27$, with full communication, none of the inter-agent messages arrive in time while reduced communication still converge to small value-function values, see~\cref{fig:embfour}.
However, LSTM+FFLH receives around $8$ percentage points less information on average compared to SHRED (\cref{tab:embeddedperformance}) although both communicate the same amount of information and use identical packet sizes.
A similar result can be observed when increasing the number of agents to eight.
Here, the full communication already collapses at $\bar p = 12$ due to more demanding communication conditions while SHRED continues to converge to small value function values, see both~\cref{tab:embeddedperformance} and~\cref{fig:embeddedexemplarytaskeight}, whereas LSTM+FFLH receives less information than in the scenario with four robots and fails to converge.
This performance decline is likely due to the computational overhead of the decoder LSTM, requiring about six times the amount of time compared to the FNN decoder used for SHRED per robot leading also to a decline in received state updates under challenging conditions.
Increasing $H$ to $25$, which SHRED cannot handle without retraining, both communication without reduction and LSTM+FFLH-based reduction struggle already at $\bar p = 10$.
Although LSTM+FFLH received over $90\%$ of information, it struggles to converge while, interestingly, without communication reduction, the formation managed to converge sometimes even when receiving less than $40\%$ of the communicated information.
Under these conditions, LSTM+FFLH receives substantially less of the state updates in time.
This is likely due to the increased computational overhead of the decoder resulting in the optimization using some of the time reserved for state updates.
\begin{figure}[btp]
    \centering
     \includegraphics{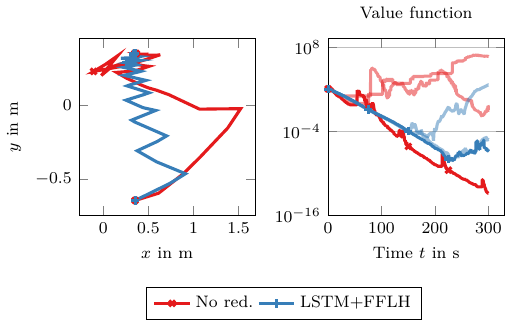}
    \caption{Left: Trajectories for one robot of an example scenario with $H=25$ and $\bar p = 10$ using no reduction (No red.) in the communication and LSTM+FFLH communication reduction. Right: The value functions of one of the robots for all conducted experiment repetitions.}
    \label{fig:embeddedexemplarytask0717}
\end{figure}%

\section{Conclusion}\label{sec:conclusion}
\noindent This work proposed and analyzed different neural network architectures to construct semantic encoding for data transmission based on LSTMs to reduce the inter-agent communication in distributed model predictive control.
Using SHRED to reduce the inter-agent communication resulted in convergence performance closely matching the performance of full communication in simulation and has been shown to perform well in experiments with physical wireless communication even in conditions where full communication faltered.
Additionally, SHRED outperforms the autoencoder-based communication reduction from~\cite{SchizEbel25}.
However, SHRED uses a standard FNN in the decoder paired with an LSTM encoder, such that retraining is required as soon as the prediction horizon length is changed.
Although the training process may be automated as it proved benign, it introduces a burden.

Using an LSTM also in the decoder allows to train a single model for multiple prediction horizons.
This generalization affected specialization such that none of the models managed to match the performances of SHRED.
Moreover, an encoder-decoder LSTM also manages to converge under some conditions where the communication without reduction is overwhelmed.
However, the larger the formation or the longer the prediction horizon, the larger the computational overhead is in the decoder counteracting the benefits of the communication reduction and shifting the bottleneck from communication back to computation.
In these experiments, it was seen that DMPC can be surprisingly robust, sometimes converging when only receiving about a third of the amount of communicated information.
Whether and when there is inherent robustness to lossy or intermittent communication in DMPC could be worth further investigation from a theoretical standpoint, compare the recent work~\cite{ShenEtAl26} investigating issues like this for a different setup. 
In a next step, the methods should be tested in a setup including the additional challenge that moving communication endpoints introduce.
Additionally, the internal structure of the OCP solutions along the prediction horizon may be considered more specifically to guide training.

\section{Acknowledgments}
\noindent This work was partly supported by (a) the Research Council of Finland through (i) ECO-NEWS (No.~358928), (ii) X-SDEN (No.~349965), and  (iii) OptAd (No.~372723); (b) EU MSCA COALESCE project (No.~101130739).

\end{document}